\documentclass{article}
\usepackage[english]{babel}
\usepackage{here}
\usepackage{cite}
\usepackage[a4paper,top=2cm,bottom=2cm,left=3cm,right=3cm,marginparwidth=1.75cm]{geometry}

\usepackage{amsmath}
\usepackage{graphicx}
\usepackage[colorlinks=true, allcolors=blue]{hyperref}
\usepackage{comment}
\usepackage{wrapfig}
\usepackage{caption}
\usepackage[normalem]{ulem}
\usepackage{gensymb}
\usepackage{amssymb}
\usepackage{array}
\usepackage{lineno}
\usepackage{multirow}
\usepackage{multicol}
\usepackage{authblk}

\title{\textbf{First operation of LArTPC in the stratosphere as an engineering GRAMS balloon flight (eGRAMS)}} 

\author[1]{R.~Nakajima}
\author[2]{S.~Arai}
\author[1]{K.~Aoyama}
\author[1]{Y.~Utsumi}
\author[3]{T.~Tamba}
\author[4]{H.~Odaka}
\author[1]{M.~Tanaka}
\author[1]{K.~Yorita}
\author[1]{S.~Arai}
\author[5]{T.~Aramaki}
\author[6]{J.~Asaadi}
\author[2]{A.~Bamba}
\author[7]{N.~Cannady}
\author[8]{P.~Coppi}
\author[7]{G.~De Nolfo}
\author[9]{M.~Errando}
\author[10]{L.~Fabris}
\author[4]{T.~Fujiwara}
\author[11]{Y.~Fukazawa}
\author[7]{P.~Ghosh}
\author[2]{K.~Hagino}
\author[4]{T.~Hakamata}
\author[1]{U.~Hijikata}
\author[12]{N.~Hiroshima}
\author[2]{M.~Ichihashi}
\author[13]{Y.~Ichinohe}
\author[4]{Y.~Inoue}
\author[1]{K.~Ishikawa}
\author[4]{K.~Ishiwata}
\author[2]{T.~Iwata}
\author[14]{G.~Karagiorgi}
\author[2]{T.~Kato}
\author[4]{H.~Kawamura}
\author[7]{J.~Krizmanic}
\author[5]{J.~Leyva}
\author[14]{A.~Malige}
\author[7]{J.G.~Mitchell}
\author[7]{J.W.~Mitchell}
\author[15]{R.~Mukherjee}
\author[16]{K.~Nakazawa}
\author[16]{K.~Okuma}
\author[14]{K.~Perez}
\author[5]{N.~Poudyal}
\author[14]{I.~Safa}
\author[7]{M.~Sasaki}
\author[14]{W.~Seligman}
\author[4]{K.~Shirahama}
\author[17]{T.~Shiraishi}
\author[18]{S.~Smith}
\author[11]{Y.~Suda}
\author[5]{A.~Suraj}
\author[11]{H.~Takahashi}
\author[2]{S.~Takashima}
\author[14]{S.~Tandon}
\author[4]{R.~Tatsumi}
\author[19]{J.~Tomsick}
\author[17]{N.~Tsuji}
\author[20]{Y.~Uchida}
\author[18]{S.~Watanabe}
\author[1]{Y.~Yano}
\author[21]{K.~Yawata}
\author[22]{H.~Yoneda}
\author[4]{M.~Yoshimoto}
\author[5]{J.~Zeng}

\affil[1]{Waseda University, 3-4-1 Okubo, Shinjuku-ku, Tokyo 169-8555, Japan}
\affil[2]{University of Tokyo, 7-3-1, Hongo, Bunkyo-ku, Tokyo 113-8654, Japan}
\affil[3]{JAXA, 3-1-1 Yoshinodai, Chuo-ku, Sagamihara City, Kanagawa 252-5210, Japan}
\affil[4]{Osaka University, 1-1 Machikaneyama-cho, Toyonaka, Osaka 560-0043, Japan}
\affil[5]{Northeastern University, 360 Huntington Avenue, Boston, MA 02115, USA}
\affil[6]{University Texas Arlington, 701 South Nedderman Drive, Arlington, TX 76019, USA}
\affil[7]{NASA GSFC, 8800 Greenbelt Road, Greenbelt, MD 20771, USA}
\affil[8]{Yale University, P.O. Box 208101 New Haven, CT 06520-8101, USA}
\affil[9]{Washington University at St. Louis, One Brookings Drive, St. Louis, MO 63130-4899, USA}
\affil[10]{Oak Ridge National Laboratory, 5200, 1 Bethel Valley Rd, Oak Ridge, TN 37830, USA}
\affil[11]{Hiroshima University, 1-3-2, Kagamiyama, Higashi Hiroshima-shi, Hiroshima 739-0046, Japan}
\affil[12]{Yokohama National University, Yokohama 240-8501, Japan}
\affil[13]{RIKEN, Hirosawa 2-1, Wako-shi, Saitama 351-01, Japan}
\affil[14]{Columbia University, New York, NY 10027, USA}
\affil[15]{Barnard College, 3009 Broadway, New York, NY 10027, USA}
\affil[16]{Nagoya University, Furo-cho, Chikusa-ku, Nagoya, Aichi 464-8601, Japan}
\affil[17]{Kanagawa University, 3-27-1, Rokkakubashi, Kanagawa-ku, Yokohama-shi, Kanagawa 221-0802, Japan}
\affil[18]{Howard University, 2400 6th St NW, Washington, DC 20059, USA}
\affil[19]{University of California Berkeley, University Avenue and, Oxford St, Berkeley, CA 94720, USA}
\affil[20]{Tokyo University of Science, 2641 Yamazaki, Noda, Chiba 278-8510, Japan}
\affil[21]{National Defense Medical College, 3-2 Namiki, Tokorozawa, Saitama 359-8513, Japan}
\affil[22]{Universität Würzburg, Emil-Fischer-Str. 31, D-97074 Würzburg, Germany, Sanderring 2, 97070 W\\"{u}rzburg, Germany}

\begin{document} 
\maketitle 

\begin{abstract}
GRAMS (Gamma-Ray and AntiMatter Survey) is a next-generation balloon/satellite experiment utilizing a LArTPC (Liquid Argon Time Projection Chamber), to simultaneously target astrophysical observations of cosmic MeV gamma-rays and conduct an indirect dark matter search using antimatter. While LArTPCs are widely used in particle physics experiments, they have never been operated at balloon altitudes. An engineering balloon flight with a small-scale LArTPC (eGRAMS) was conducted on July 27th, 2023, to establish a system for safely operating a LArTPC at balloon altitudes and to obtain cosmic-ray data from the LArTPC. The flight was launched from the Japan Aerospace Exploration Agency's (JAXA) Taiki Aerospace Research Field in Hokkaido, Japan. The total flight duration was 3 hours and 12 minutes, including a level flight of 44 minutes at a maximum altitude of 28.9~km. The flight system was landed on the sea and successfully recovered. The LArTPC was successfully operated throughout the flight, and about 0.5 million events of the cosmic-ray data including muons, protons, and Compton scattering gamma-ray candidates, were collected. This pioneering flight demonstrates the feasibility of operating a LArTPC in high-altitude environments, paving the way for future GRAMS missions and advancing our capabilities in MeV gamma-ray astronomy and dark matter research.
\end{abstract}

\clearpage 

\section{Introduction} \label{sec1}%

\subsection{MeV Gamma-Ray Astronomy}
The astrophysical observations at MeV energies have not yet been well-explored, the so-called MeV-gap. However, it is a missing key for understanding the multi-messenger astrophysical phenomena, the origin of heavy elements, and cosmic-ray acceleration \cite{gamma}. The dominant reactions for the MeV gamma rays are Compton scattering and photo-absorption below 10~MeV. A Compton camera is commonly used to determine the direction of MeV gamma rays \cite{DOGAN1990501}. Because of the nature of the gamma-ray interactions in this energy range and the reconstruction techniques required, a sufficiently large detector is needed to contain the gamma rays, in addition to fine-grained spatial readout and excellent spatial and energy resolution. Thus, the sensitivity of current gamma-ray missions for the MeV energy range is not as good as that of detectors for the other energy bands. COMPTEL (The Imaging COMPton TELescope) aboard the Compton Gamma-Ray Observatory, launched in 1991, succeeded in creating the first all-sky map and detecting some nuclear gamma-ray lines, but it could only detect about 30 sources for steady objects in the range of 0.75 MeV to 30 MeV \cite{Schonfelder00}. The INTEGRAL mission, launched in 2002, improved the spectra of various nuclear gamma-ray lines and maps, particularly in the lower energy range, up to 8 MeV~\cite{INTEGRAL}. 

\subsection{Cosmic-Ray Antiparticles}
Various cosmic particles, including antiparticles such as antiprotons, are incident on the Earth from space. The fluxes of antiprotons have been detected and measured by numerous experiments \cite{AMS02,BESS,PAMELA}. The majority of the measured antiproton flux is explained by secondary production in which primary cosmic rays such as protons colliding with interstellar gas: $p_{\mathrm{CR}}+p_{\mathrm{ISM}} \rightarrow \bar p + X$. However, a potential excess exists in the low energy region around 1~GeV. While still debated, this excess could be potentially described by the annihilation of dark matter (DM) with a mass around 50 GeV \cite{AMS02_pbarexcess}. Despite the high-statistics antiproton measurements, cosmic-ray antideuterons have never been observed. Currently, the only upper limit on the antideuteron flux is placed by BESS Polar-II experiment \cite{BESSlimit}. When considering the dark matter model that describes the potential antiproton excess, the predicted antideuteron flux at energies below 0.2 GeV/n from dark matter annihilation is significantly larger (S/N $>$ 100) than the expected astrophysical secondary flux. This difference arises due to the kinematic limitations on the secondary production. In contrast, dark matter annihilation lacks these limitations, allowing the production of low-energy antideuterons. Consequently, the search for antideuterons has gained attention as a potential smoking gun for indirect DM search. 


\subsection{The GRAMS Experiment} \label{sec1.3}
Liquid Argon Time Projection Chambers (LArTPCs) have become standard technology in particle physics experiments, particularly for direct dark matter detection and neutrino experiments \cite{darkside20k,DUNE,MicroBooNE}. This is due to liquid Argon's favorable properties: dense (40\% denser than water), abundant (gas argon makes 1\% of the atmosphere), and highly sensitive to incoming particles (producing 40 photons/keV for scintillation and 1 fc/mm for ionization). However, LAr is cryogenic and the difference between the melting and boiling points is only 3.5$^\circ$C at 1~atm pressure. Furthermore, LArTPCs require high-purity liquid argon ($<$ 1 ppb impurity) to function effectively. Therefore, infrastructure to maintain and control LAr is essential for LArTPC experiments. 

\begin{figure}[htbp]
  \begin{center}
    \includegraphics[height=7cm]{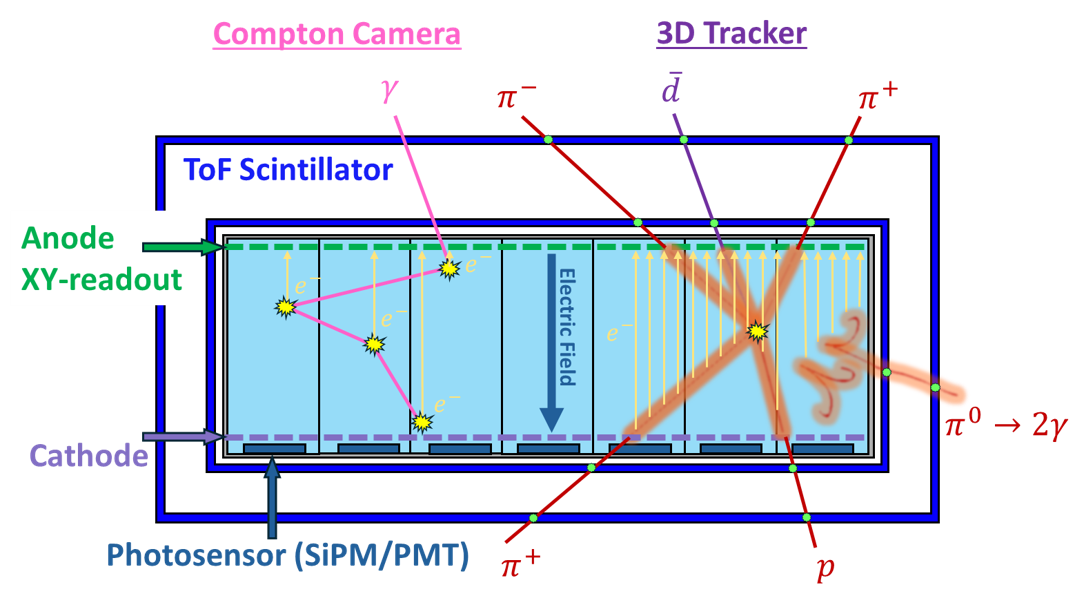}
  \end{center}
  \vspace{-0.4 cm}
  \caption[GRAMS Detector]
  {GRAMS LArTPC conceptual design. (Left side) Multiple Compton scattering of MeV gamma-ray. (Right side) Antideuteron capture into annihilation with argon nucleus producing multiple hadrons\cite{GRAMS_First}.}
  \label{fig:GRAMSLArTPC}
\end{figure}

The GRAMS (Gamma-Ray and AntiMatter Survey) experiment is a pioneering next-generation balloon/satellite experiment. It is the first project to utilize a LArTPC for simultaneous astrophysical observations of cosmic MeV gamma-rays and indirect dark matter searches using cosmic antiparticles \cite{GRAMS_First}. Figure \ref{fig:GRAMSLArTPC} illustrates the conceptual design of GRAMS and the detection principle of a LArTPC. As shown in Figure \ref{fig:GRAMSLArTPC}, the GRAMS detector consists of a LArTPC surrounded by two layers of Time-of-Flight (ToF) scintillators. For detecting charged particles, the ToFs can provide velocity information of the incoming particle. Meanwhile, vetoing charged particles can provide time windows for detecting gamma rays. When a charged particle enters the LArTPC, energy is deposited in the form of scintillation light and ionization electrons. The wavelength spectrum of the scintillation light peaks at 128~nm and therefore needs to be converted into visible light for detection by high-sensitivity cryogenic photosensors (PMTs or SiPMs). The ionization electrons drift toward the anode due to the applied electric field. The X and Y positions are determined at the anode (with a few mm pitch) while the Z positions are determined from the drift time of the ionization electrons after the scintillation light is measured. It is important to note that electronegative impurities such as water and oxygen absorb ionization electrons and therefore high-purity LAr is needed to achieve high sensitivity across the entire detection area. 

As shown in the left side of Fig.~\ref{fig:GRAMSLArTPC}, when a gamma-ray enters the LArTPC, it may undergo several Compton scatterings before being photo-absorbed or escape the LArTPC, allowing the detector to function as a Compton camera with a large effective area. The energy and position of the scattering points in the LArTPC are used in reconstruction algorithms to determine the direction of the incident gamma ray. Therefore, the LArTPC works as a Compton camera with a large effective area. \cite{Yoneda, Takashima}. 

As shown in the right side of Fig.~\ref{fig:GRAMSLArTPC}, when a charged antiparticle ($\bar{p}, \bar{d}, \mathrm{\overline{He}} $) slows down and stops in LAr through ionization energy loss, it forms an exotic atom with the argon nucleus. The exotic atom de-excites, releasing X-rays, and at the end of the transition, the antiparticle annihilates with the nucleus, releasing multiple hadrons, which can also be detected \cite{ARAMAKI20166}. In summary, the LArTPC can provide 3D-tracking information with deposited energy along the trajectory (dE/dx) for charged particles, allowing for efficient particle detection and the detection of annihilation products (if any). Combined with velocity information from the ToFs, this enables strong particle identification.

The GRAMS project has three major milestones. Firstly, the operation of LArTPC as a Compton camera. The R\&D for improving the position and energy resolution for a LArTPC to accurately detect Compton scatterings is required. Secondly, a thorough understanding of the reaction of antiparticles in LAr is required~\cite{T98}. Although the antiparticle capture reaction has been observed and studied with various target nuclei \cite{antiprotoncapture}, high statistics measurements of antiparticle capture in LAr have not been conducted yet. Thirdly, the operation of LArTPC at balloon altitudes serves as a pivotal milestone for future GRAMS flights. Although LArTPCs have been widely used for underground experiments, they have never been operated at balloon altitudes. The only other liquid noble gas TPC operated at balloon altitudes is the Liquid Xenon Gamma-Ray Imaging Telescope (LXeGRIT) experiment which used liquid xenon TPC as a Compton camera to search for MeV gamma rays \cite{LXeGRIT}. To establish the milestone of operating a LArTPC at balloon altitudes, an engineering balloon flight with a small compact LArTPC (eGRAMS) was conducted on 27th July 2023, where a safe and robust operation of LArTPC at balloon altitude conditions was established.

This paper will give an overview and discuss the results of the engineering flight as a part of the GRAMS project. Section 2 provides an overview of this engineering flight. The LAr handling system and the LArTPC used in this flight will be described in Section 3. A timeline and summary of the flight will be presented in Section 4. The results of the flight including the data from the LArTPC will be explained in Section 5. A final summary and the upcoming events for GRAMS will be described in Section 6. 


\section{The B23-06 Flight Campaign} \label{sec2} 

\begin{figure}[htbp]
\centering
\begin{minipage}[b]{0.5\linewidth}
\centering
\includegraphics[width=7.5cm]{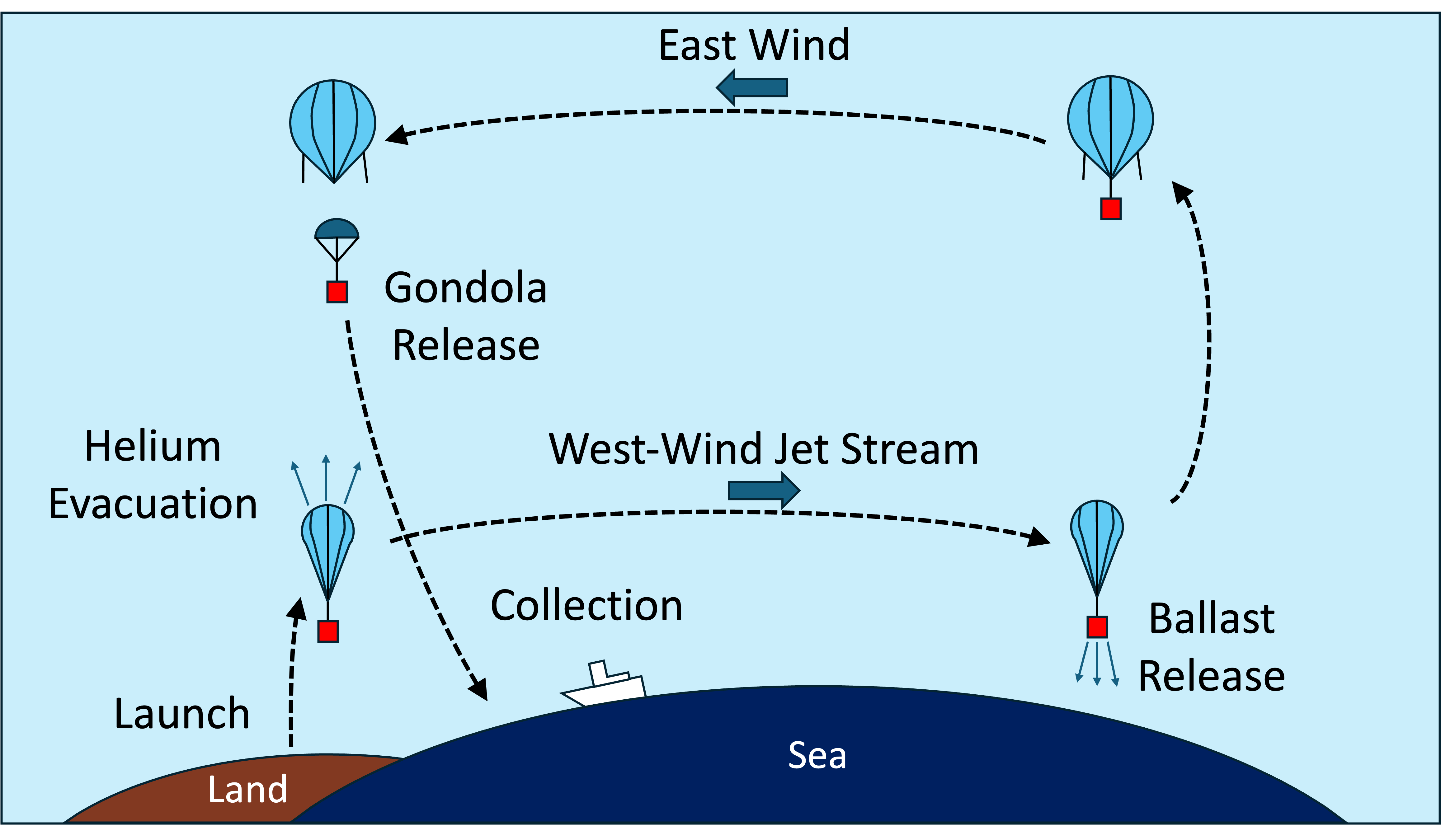}
\caption{``Boomerang'' flight trajectory from TARF.}
\label{fig:boomerang}
\end{minipage}
\hspace{0.5cm}
\begin{minipage}[b]{0.4\linewidth}
\centering
\includegraphics[width=6cm]{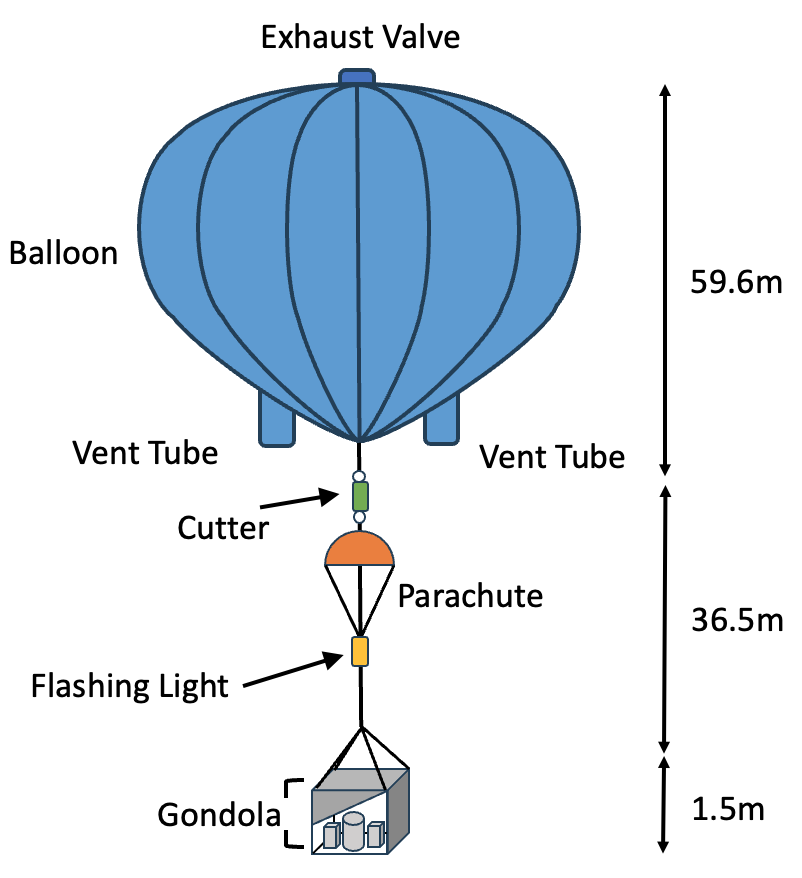}
\caption{A schematic diagram of the B23-06 balloon system.}
\label{fig:eGRAMSsystem}
\end{minipage}
\end{figure}

Since 2009, science flights that were selected annually by the ISAS/JAXA scientific balloon program advisory committee have been conducted at Taiki Aerospace Research Field (TARF) located in Taiki, Hokkaido, Japan \cite{TARF}. On September 30, 2022, we submitted a proposal for the GRAMS engineering flight as a JAXA domestic scientific balloon program (PI: Hirokazu Odaka)~\cite{eGRAMS}. In April of 2023, this engineering flight was officially approved as one of the balloon flights for 2023 with two piggyback experiments, designated as B23-06. Flight trajectory for balloons launched from TARF is shown in Figure \ref{fig:boomerang}. During the flight season which spans from May to September, the seasonal wind pattern is east wind in the upper stratosphere and strong west-wind jet streams at lower altitudes. This results in a ``boomerang'' flight, increasing the flight duration and enabling the collection of the payload and the balloon close to the land. 

The whole flight system consisting of a balloon, parachute, and gondola is shown schematically in Figure~\ref{fig:eGRAMSsystem}. The scientific equipment (LArTPC), ballast, and bus system are contained in the gondola. The ballast is used to control the gondola's weight, which enables the control of the balloon's altitude. The bus system is used for telemetry and command communications with the ground. The gondola is connected to the parachute and balloon with a nylon rope which can be cut using a cutter that is initialized by telemetry command. The balloon has an exhaust valve that can release the helium gas, to control the balloon's altitude. In addition, vent tubes are used to equalize the pressure inside the balloon to the atmospheric pressure. The total weight of the system was 643 kg. Table \ref{tab:weight} shows the weights of the individual components.


\begin{table}
\caption{Weight of each component for the B23-06 flight.}
\label{tab:weight}
\centering
\begin{tabular}{l|r}
\hline
\textbf{Components}                  & \textbf{Weight (kg)} \\ \hline
\textbf{Balloon Subtotal}              &   \textbf{201}     \\ 
\ \ \ \ Balloon                      &   152      \\ 
\ \ \ \ Packing                      &   49       \\ \hline
\textbf{Gondola Subtotal}                     &  \textbf{442}     \\ 
\ \ \ \ Gondola Frame                        &  75       \\ 
\ \ \ \ Ballast                      &  158      \\ 
\ \ \ \ GRAMS Cryostat               &  71                   \\ 
\ \ \ \ Liquid Argon                 &  15                   \\ 
\ \ \ \ GRAMS Pressurized Vessel     &  31                   \\ 
\ \ \ \ Others (piggybacks, controller, etc)                      &  92     \\ \hline 
\textbf{Total}                       & \textbf{643}     \\ \hline
\end{tabular}
\end{table}

The requested flight parameters were maximum altitude of $\geq$ 25~km and level flight duration $\geq$ 0~min. The flight parameters were set to maximize the chances of flight. The minimum success for B23-06 was defined as maintaining and monitoring the temperature and pressure of the vessel containing LAr and operating the LArTPC during the ascend. Furthermore, the full success was to operate the LArTPC to obtain cosmic-ray and atmospheric gamma-ray data during a level flight. 

\begin{figure}[htbp]
\centering
\begin{minipage}[t]{0.45\linewidth}
\centering
\includegraphics[width=5cm]{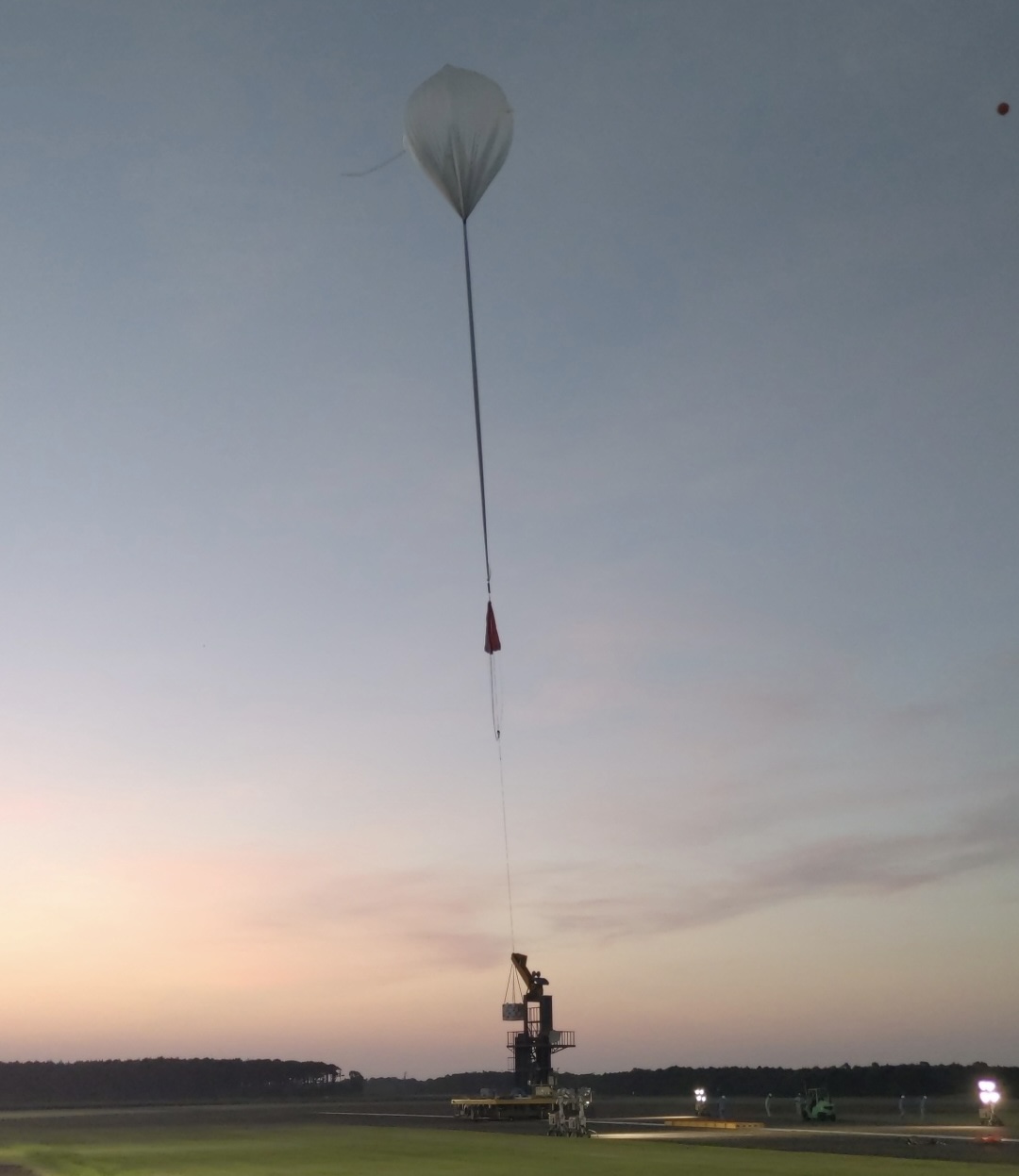}
\caption{B23-06 launch at 3:55~AM.}
\label{fig:launch}
\end{minipage}
\hspace{0.5cm}
\begin{minipage}[t]{0.45\linewidth}
\centering
\includegraphics[width=5cm]{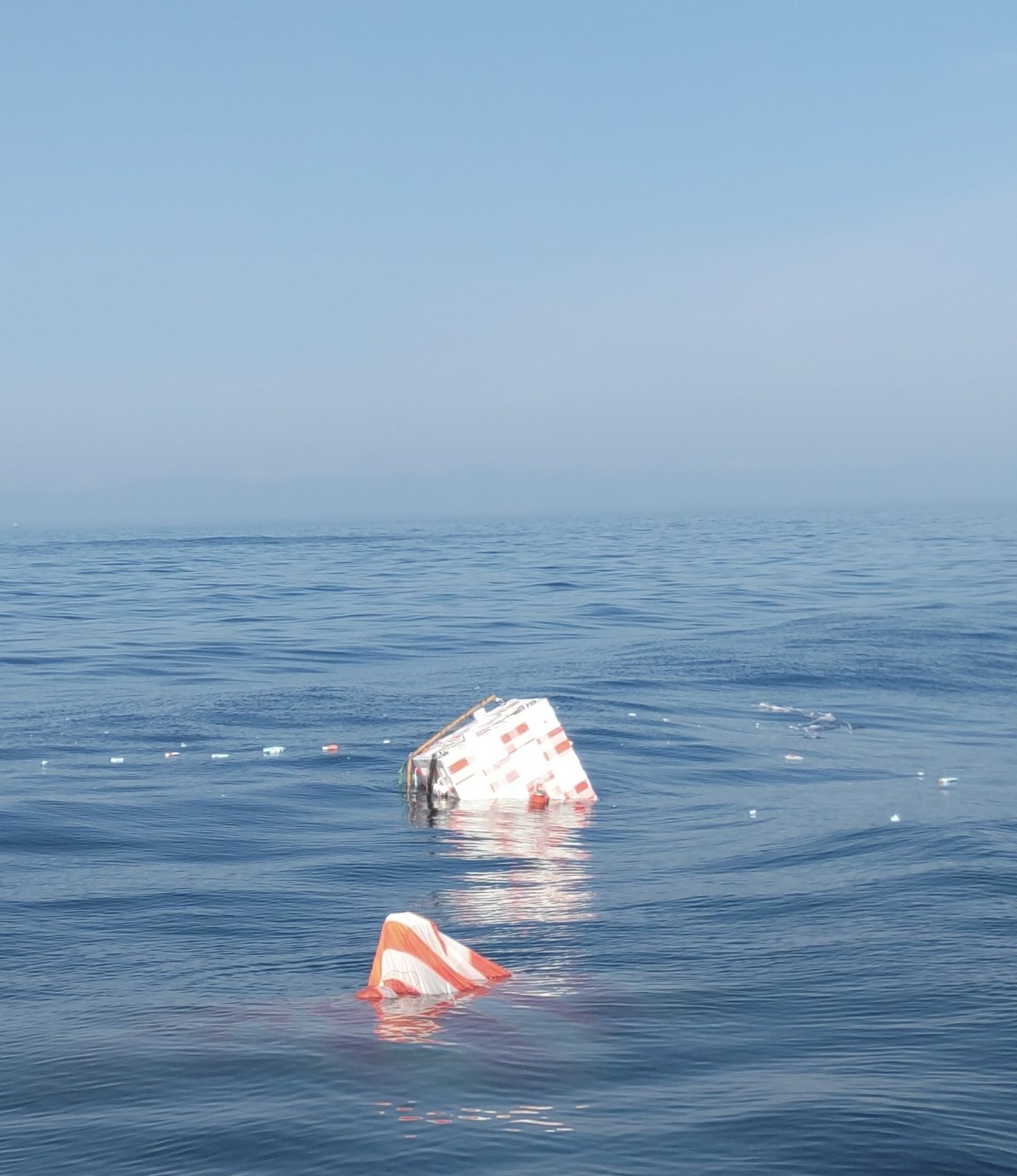}
\caption{Gondola floating after landing on the sea.}
\label{fig:landing}
\end{minipage}
\end{figure}

The R\&D phase for this flight was approximately 6 months, following its proposal in November 2022. In the initial phase (Dec 2022 to May 2023) which was held mostly at Waseda University, the hardware components were designed and manufactured. Before transporting the setup to JAXA's Sagamihara campus, a LArTPC test was completed with the integration of the readout electronics used for the flight. The next phase (May to June 2023) of preparation was conducted at JAXA's Sagamihara campus where the software was mainly developed. Furthermore, a vacuum thermostatic chamber was used to test if the detector system was operational under balloon altitude conditions. Finally, in the final phase (June to July 2023), the gondola setup was transported to TARF where full setup operational tests and telemetry configuration tests took place. The Flight Readiness Review (FRR) which determines if the whole system is ready for launch was conducted on July 5th. From the day of the FRR to the day of launch, the cryostat was continuously vacuum-pumped to minimize outgassing, and sufficient LAr for the flight was secured every day. On the day of the flight (27th July 2023), the final operational test took place at 12:00~AM. Upon clearance, flight B23-06 was launched at 3:55~AM and landed on the sea at 7:07~AM, as shown in Figure \ref{fig:launch} and Figure \ref{fig:landing}, respectively. All times mentioned hereafter will be in Japan Standard Time (JST). A detailed timeline of the day of the flight and a flight summary is given in Section \ref{sec4}.

\section{Payload Design and Components}\label{sec3}

\begin{figure}[htbp]
  \begin{center}
    \includegraphics[height=6cm]{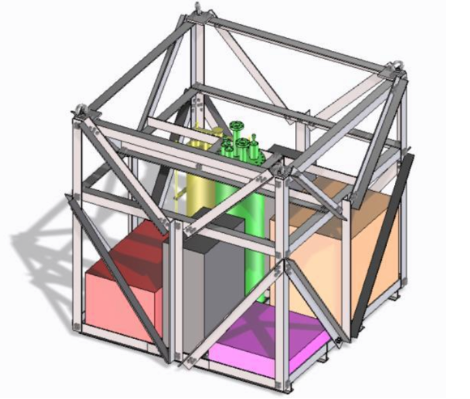}
    \includegraphics[height=7cm]{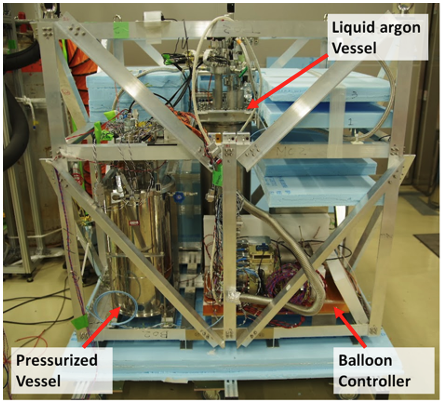}
    \includegraphics[height=6cm]{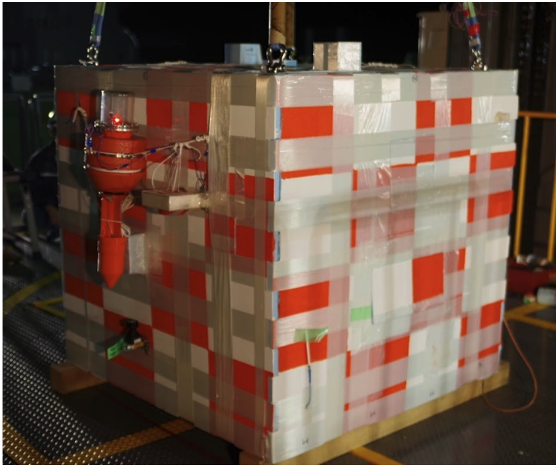}
  \end{center}
  \vspace{-0.4 cm}
  \caption[Images of payload]
  {(Top-left) Layout of each component in the gondola. The LAr vessel (green), the pressurized vessel (yellow), the bus communication system (orange), and two piggyback experiments (red and purple). (Top-right) Photo of the cryostat and pressurized vessel in the gondola. (Bottom) The exterior of the gondola right before the launch, showing the heat insulator.}
  \label{fig:GondolaBall}
\end{figure}

As shown in Figure \ref{fig:GondolaBall} (top-left), the LAr vessel (shown in green) is placed in the center of the gondola with two ballast boxes on its side. The pressurized vessel (shown in yellow) containing the electronics is placed in one corner. In the other three corners of the gondola, the bus system (orange) and two piggyback experiments (red and purple) are placed. As shown in Figure~\ref{fig:GondolaBall} (bottom), the entire gondola is covered with 25~mm (bottom half) and 75~mm (top half) heat insulation material (Styrofoam B2) as the atmospheric temperature drops to at least $-$70$^\circ$C at higher altitudes. The Styrofoam on the outer layer of the gondola was colored with orange and white paint to improve visibility. As the Styrofoam also acts as a float after landing on the sea, extra Styrofoam was inserted in the gondola (top-right photo of Figure~\ref{fig:GondolaBall}). 

The gondola was designed to withstand 3~g in horizontal (2 axes) and vertical directions for the impact during launch and 7.5~g in the vertical direction for the impact during parachuting with 50\% of the safety margin. The gondola frame mainly consisted of L-shaped aluminum (A6065-t5, $\rm 500\times500\times t5~mm^3$). The overall size is a cube of $\rm 1.2\times1.2\times1.2~m^3$. The frames are fastened to each other with stainless steel screws. Stainless steel eye-nuts were used at the suspension points, and extra-strong duralumin (A7075) was installed at the eye-nut fixing points.

\subsection{LAr Handling System}\label{sec3.1}
\begin{figure}[htbp]
  \begin{center}
    \includegraphics[height=8.5cm]{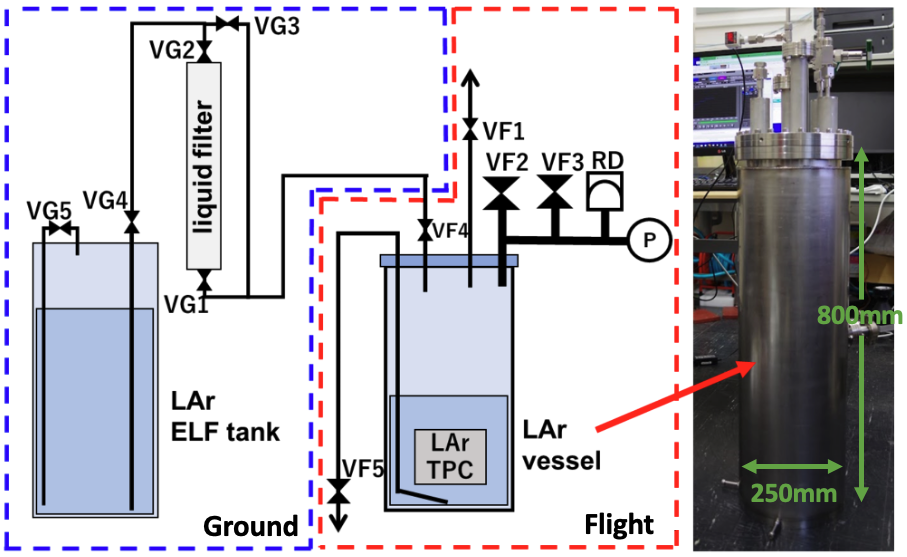}
  \end{center}
  \vspace{-0.4 cm}
  \caption[Schematic diagram of LAr operation system and an image of LAr vessel]
  {Schematic diagram of LAr operation system and an image of LAr vessel. VG: Valves used in the ground system. VF: Valves used in the flight system.}
  \label{fig:LArHandSys}
\end{figure}

A schematic diagram of the LAr handling system and a picture of the LAr vessel are shown in Figure \ref{fig:LArHandSys}. The LAr is held in a stainless steel vacuum-insulated vessel 80~cm high and 25~cm in diameter. As electronegative impurities, such as oxygen and water, in LAr absorb ionization electrons, high purity (less than 1 ppb of impurity) is required in operating a LArTPC. To ensure high-purity argon was filled, the LAr vessel was vacuum-pumped to at least 10$^{-3}$ Pa before filling through the manual valve (VF3 in Figure \ref{fig:LArHandSys}). Then, LAr was filled through a handmade filter, which consists of a molecular sieve and reduced copper~\cite{CURIONI2009306}. Sufficient LAr was filled at 8:00 am on the day before the launch. Then, the pressure and temperature of the cryostat and the LAr liquid level were continuously monitored until the launch.

To maintain argon in its liquid state, it was necessary to keep the inner pressure and temperature above argon's triple point (0.7~bar and 84~K). As the top flange will be near room temperature, to reduce the heat inflow as low as possible and prevent LAr from contacting the top flange causing a sudden increase in pressure, the LAr vessel was designed to be thin (2 mm) and tall (800 mm). Thus, this vessel has a heat inflow rate of around 15~W on the ground (LAr evaporation rate of 0.25 L/hour and LAr level reduction of 7~mm/hr in the vessel) and can hold enough LAr to operate the detector for more than 24~hours. The vessel pressure was maintained by evacuating the evaporated argon (approximately 3~L/min) using an absolute pressure valve manufactured by TAVCO (VF1 in Figure \ref{fig:LArHandSys}). 
The operating pressure is 17.5~PSI (1.2~atm), and the maximum flow rate is about 20~L/min. Under normal operation, the pressure can be controlled through the absolute pressure valve alone. The pressure inside the detector was measured by an absolute pressure gauge, and the temperature inside the vessel was measured by a platinum resistance thermometer. If the pressure rises sharply due to the impact of launch etc., a differential pressure valve (SL-39, Venn Corporation) with a maximum flow rate of 600~L/min is used to exhaust the gas (VF2) where the working differential pressure is $+$1.5~bar. In case of abnormal evaporation of Ar beyond the evacuation capacity of the differential pressure valve, a rupture disk (RD) with a rupture pressure of $+$2.0~bar, manufactured by V-TEX Corporation is used to exhaust the evaporated argon. The piping to all pressure valves ensured that the valve outlets would be above the sea surface after landing. In case of the gondola overturning, a water-repellent cotton ball was attached to the valve outlets. 


LAr was evacuated in the air during descent for safety recovery on the sea. A stainless steel piping was extended to the bottom of the LAr vessel and the evacuation pipe extended to the gondola frame. In this way, the difference between the inner pressure of the vessel and the atmosphere allows the LAr to be evacuated from the vessel through the siphon mechanism. The evacuation of LAr is started and stopped by a solenoid valve (J263G210LT, Nihon Asco, Ltd.) shown as VF5 in Figure \ref{fig:LArHandSys}, which can be driven at low temperature at the pipe outlet. When the evacuation is complete, the internal pressure becomes equal to the external atmospheric pressure. 


\begin{figure}[htbp]
  \begin{center}
    \includegraphics[height=6cm]{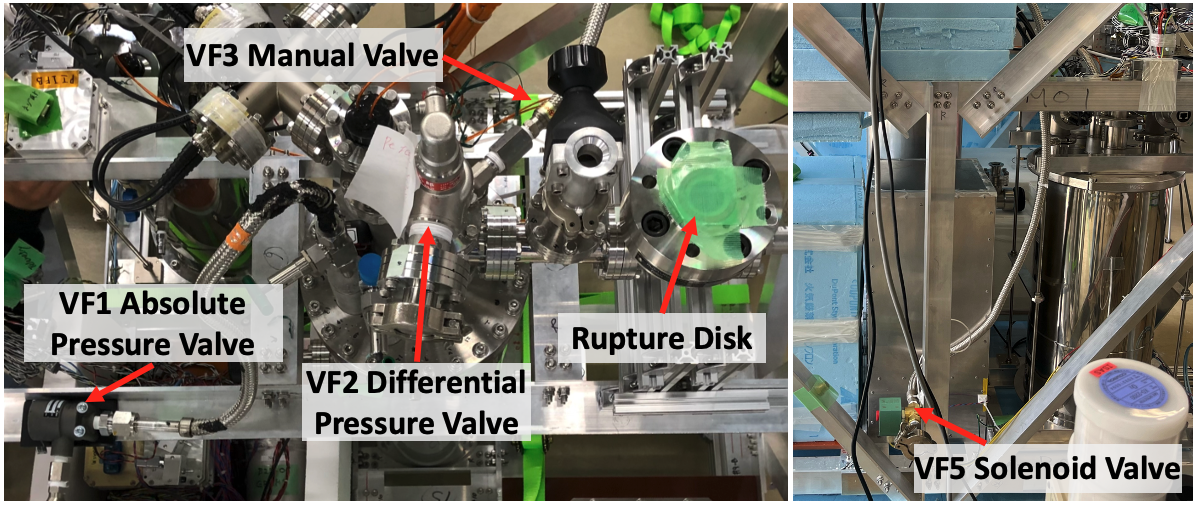}
  \end{center}
  \vspace{-0.4 cm}
  \caption[Picture of valve system on the top/side of the vessel.]
  {Image of the valves used for LAr handling in the flight system.}
  \label{fig:valves}
\end{figure}

\subsection{Detector System}\label{sec3.2}
\begin{figure}[htbp]
  \begin{center}
    \includegraphics[height=8cm]{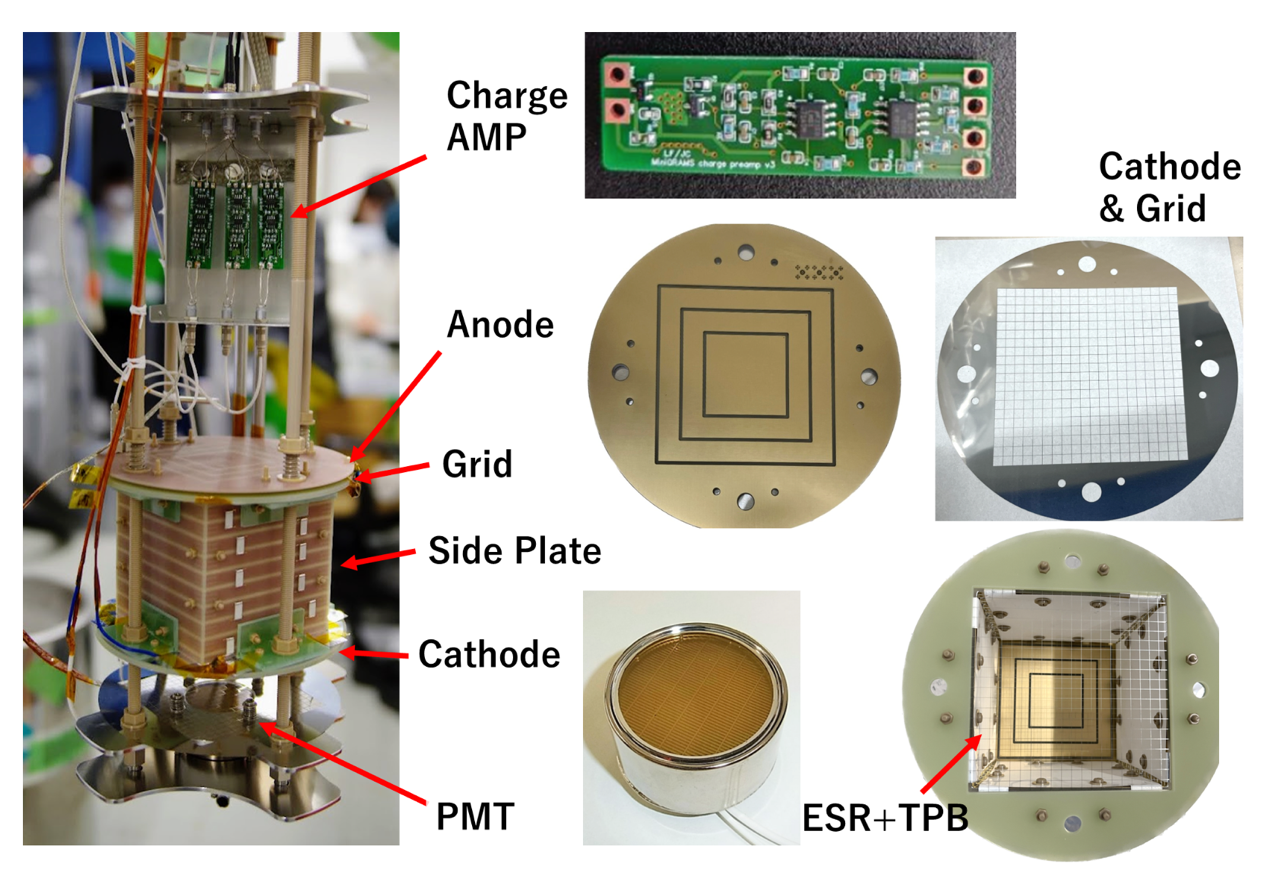}
  \end{center}
  \vspace{-0.4 cm}
  \caption[Detector System]
  {Detector system in the LAr vessel and components of the LArTPC.}
  \label{fig:LArTPC}
\end{figure}

The LArTPC consists of four side plates and an anode made with PCBs, with a stainless steel meshed grid (for the cathode and anode grid), and its active volume is $\rm 10\times10\times10~cm^3$. From the bottom to the top, the cathode is positioned at Z=0~cm, the grid at Z=10~cm, the anode at Z=10.5~cm, and the side plate is located on the side of the active volume. The electric field is formed in the vertical direction with electrodes in the side plate which have a 1~cm separation and 100~M$\Omega$~resistor in between. A 250~M$\Omega$~resistor is placed between the grid and anode (GND). During operation, a voltage of -2.5~kV was applied to the cathode, resulting in a drift electric field of 200~V/cm from the cathode to the grid and an induction field of 1~kV/cm between the grid and anode. The anode was a pad (not a wire) as a countermeasure to vibrations and impacts during flight. The anode pad is divided into three segments (channels), as shown in Figure \ref{fig:LArTPC}, with the outermost, intermediate, and innermost channels having a dimension of $\rm 9\times 9~cm^2$, $\rm 6.6\times 6.6~cm^2$, $\rm 4.2\times 4.2~cm^2$, respectively. 

The ionization electron signal is amplified and read out by a low noise charge-integration amplifier (designed based on \cite{preamp}) for each of the segmented electrodes. The gain and time constant of the charge-integrating amplifier are 2~V/pC and 500~$\mu$s (R=1~G$\Omega$, C=0.5~pF), respectively. The LAr scintillation light is wavelength-shifted to visible light (420nm) by Tetraphenyl butadiene-deposited enhanced specular reflector films on the side walls of the TPC. The scintillation light is detected by a PMT (Hamamatsu Photonics K.K. R6041-506) operated at $+$650~V located under the TPC cathode. The PMT signals were used as the start time of the charge signals from the anode as well as the trigger for the data acquisition. The baffle plates connected to the PMT and charge amplifier box are installed to ensure the TPC stays in place during any impacts or vibrations. Also, it acts as a stopper for LAr which is expected to move in the vessel during launch.

\begin{figure}[htbp]
  \begin{center}
    \includegraphics[height=9cm]{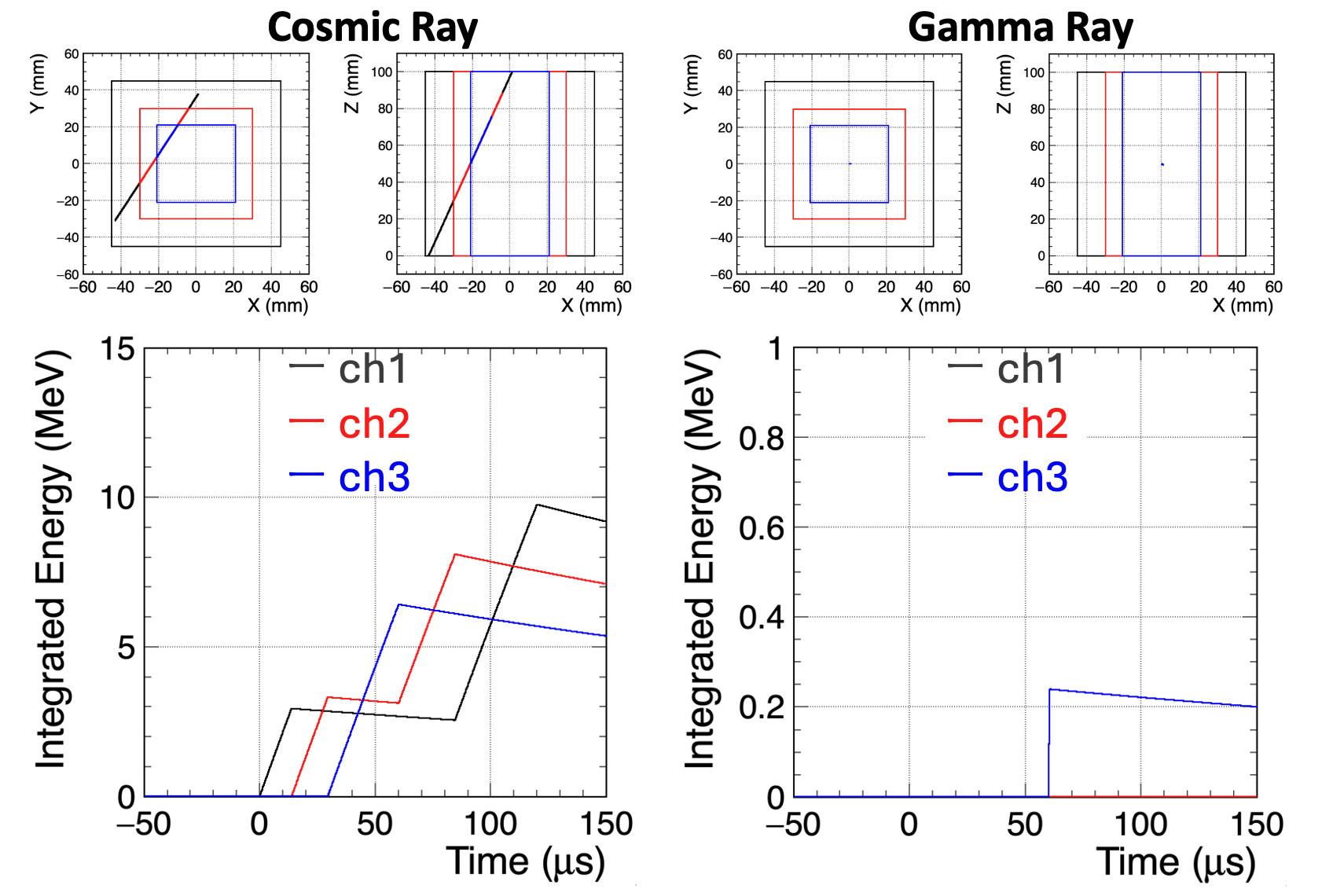}
  \end{center}
  \vspace{-0.4 cm}
  \caption[Simulated Event]
  {Simulated detector output of a cosmic ray (left) and gamma-ray (right) event. Each color represents the signal on each channel on the anode, (black) outermost channel, (blue) innermost channel, and (red) intermediate channel. }
  \label{fig:Eventsim}
\end{figure}

The simulated events of a cosmic ray penetrating through the TPC and a gamma-ray Compton scattering in the middle of the TPC are shown in Figure \ref{fig:Eventsim}. To simulate the output of the charge preamplifiers connected to each anode channel, a step simulation was conducted where a particle was generated and the energy deposited in each step along the track was converted to the output of the charge preamplifier. The outermost channel is labeled ch1, and the innermost channel as ch3. Since there is a grid below the anode, the time at which the signals on each channel start to rise represents the z position of the electron. The y-axis in the plots represents the Integrated Energy (MeV), which was obtained by converting the output of the charge preamplifier from volts to MeV using a conversion factor derived from the expected output of a Minimum Ionizing Particle (MIP) passing through the TPC. For a cosmic ray, the trajectory of the track is shown in the top two plots. For this simulated event, the cosmic ray penetrates through the TPC, hence there will be two signals on ch1 and ch2 and one signal on ch3. For the gamma-ray event, the Compton scattered event would show a point-like event, in which the signal is only present in the channel above the Compton scattering position. 

\subsection{Power Supply and Data Acquisition Systems}\label{sec3.3}
\begin{figure}[htbp]
\centering
\begin{minipage}[t]{0.47\linewidth}
\centering
\includegraphics[height=6.5cm]{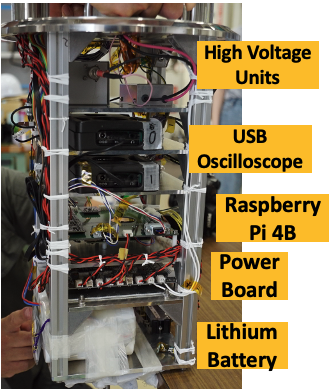}
\caption{Inner structure and layout of the pressurized vessel.}
\label{fig:PV_in}
\end{minipage}
\hspace{0.5cm}
\begin{minipage}[t]{0.47\linewidth}
\centering
\includegraphics[height=6.5cm]{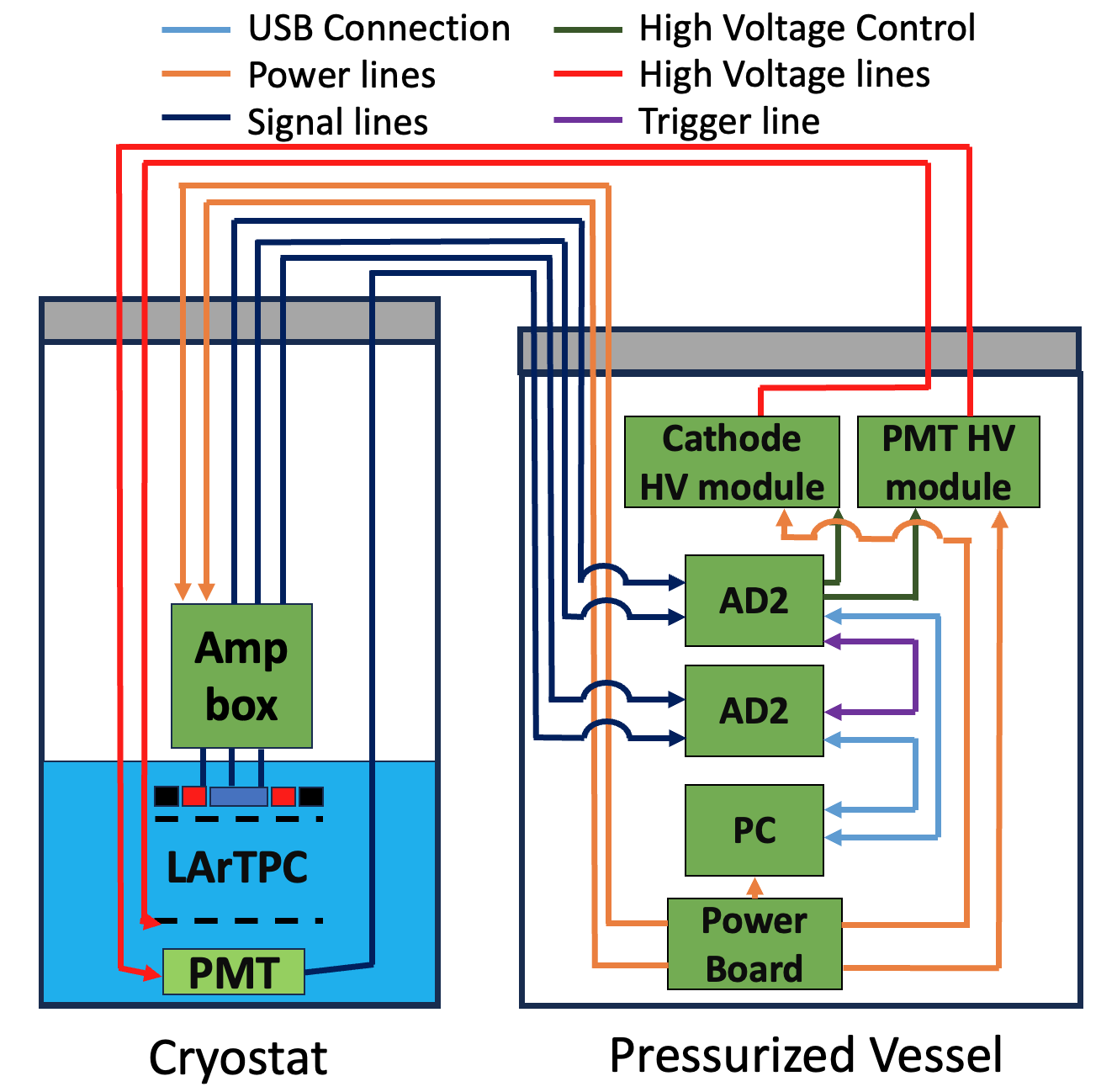}
\caption{Schematic view of the data acquisition system.}
\label{fig:DAQ}
\end{minipage}
\end{figure}

One of the technical challenges for this flight was developing the electronics under the constraints of a limited power supply, weight requirement, and balloon environments. As a countermeasure to the decreasing temperature and pressure at balloon altitudes, all the electronics were placed in a vessel with high air-tightness as shown in Figure \ref{fig:GondolaBall} (top-right). To prevent condensation inside the vessel, nitrogen gas was filled before the flight. The pressurized vessel used was the 20~L vessel of the PCN-F series manufactured by MONOVATE Co. with some customization in the top flange and the inner structure. The components inside of the vessel are shown in Figure \ref{fig:PV_in}. 

The power supply system consisted of primary lithium batteries (3B0076, Electrochem) and a custom-made power supply board (PSB, Shimafuji Electric Inc.). Lithium primary cells were used because of their large capacity and ability to be operated at low temperatures. Eight lithium primary cells were connected in series to provide 32~V (each cell has a voltage of 3.9~V, 3.0~A at room temperature). The total capacity of the battery was $30~{\mathrm{Ah}} \times 3.9~\mathrm{V} \times 8 = 936~{\mathrm{Wh}}$. The total usage of the battery after all the tests and the flight was 130~Wh (14\%). The PSB was designed to distribute the power from the battery to each module via DC-DC converters. Also, the PSB was designed to receive discrete commands (explained in the next section) through JAXA's communication system to turn on/off each module. To protect the battery that has to be used within a current of 3.0 A for safe operations, a current limiter was installed in the PSB.

As shown in Figure~\ref{fig:DAQ}, the data acquisition system consisted of a Raspberry Pi 4 Model B (RP), two units of Analog Discovery 2 (AD2, Digilent), and two high voltage modules (C14051-15 for the cathode and C13890-55 for the PMT, both made by Hamamatsu Photonics K.K.). 
The AD2 modules are connected to the RP via a USB 2.0 interface and digitized the PMT signal and LArTPC signals at a sampling rate of 50~MHz and a maximum trigger rate of 60~Hz. The high-voltage modules are controlled by two channels of digital-to-analog converters of AD2. 

The LAr environment was monitored with an absolute pressure gauge and three resistance temperature detectors (RTD). The pressure gauge output was digitized by an ADC onboarded on the PSB, and the resistance of the RTD sensors was digitized by a resistance digital converter (MAX31865, Analog Devices Co.). Both the ADC and the resistance digital converter were connected to the Raspberry Pi via its GPIO pins using SPI communication for data readout.

\subsection{Telemetry and Command System}\label{sec3.4} 
\begin{figure}[htbp]
\centering
\includegraphics[height=8cm]{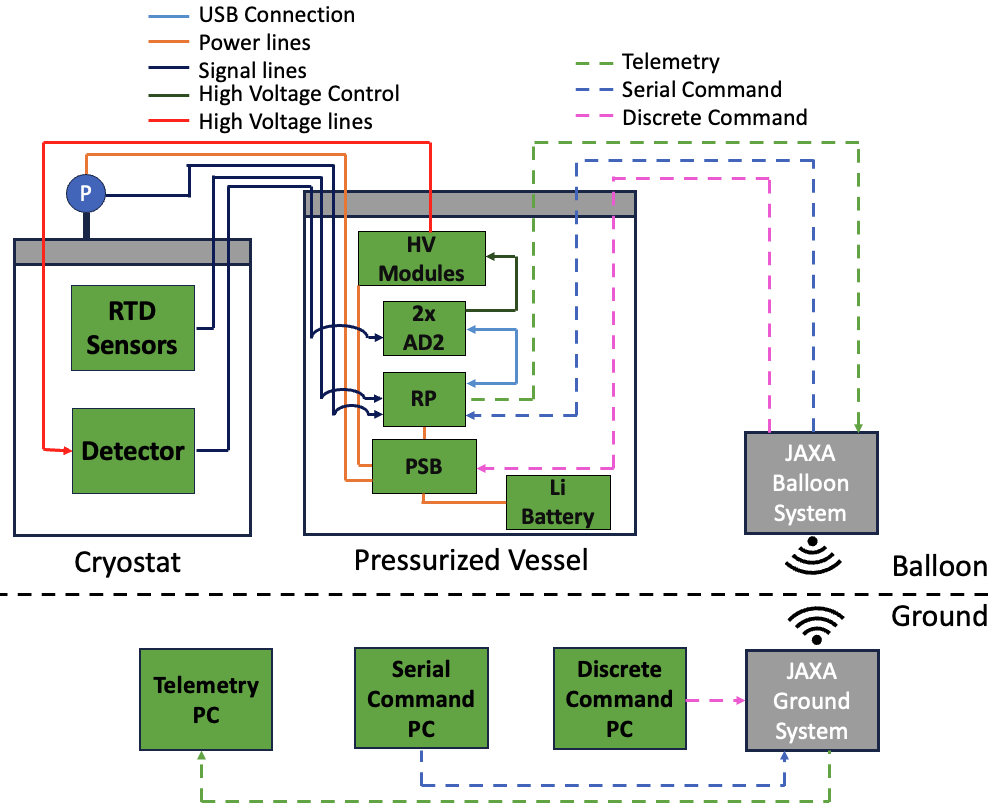}
\caption{Schematic diagram of JAXA's communication system connected to the GRAMS system.}
\label{fig:SchematicPV2}
\end{figure}

In a balloon experiment, the detector system needs to be controlled and monitored remotely. As the LArTPC requires a cryogenic operation, the temperature and pressure need to be monitored frequently and communication with the balloon was essential for a safe operation. The main methods of communication were through commands and telemetry.
There were two types of commands available for this flight: discrete commands and serial commands. The former are hardware commands used to switch on/off each device, while the latter are software commands used to send the setting information of the data acquisition and the high voltage operation. The telemetry data were continuously sent from the balloon to the ground at a rate of around 1~Hz. The communication system between the balloon and the ground system was provided by JAXA. The RP and PSB were connected to JAXA's communication equipment on the balloon as shown in Figure \ref{fig:SchematicPV2}. During the flight, various parameters such as the LAr vessel pressure were monitored through the telemetry. Also, the LArTPC waveform was checked through an on-demand telemetry, triggered by a given serial command. 


\section{Flight Summary: July 27th 2023}\label{sec4}

\begin{figure}[htbp]
  \begin{center}
    \includegraphics[height=6.0cm]{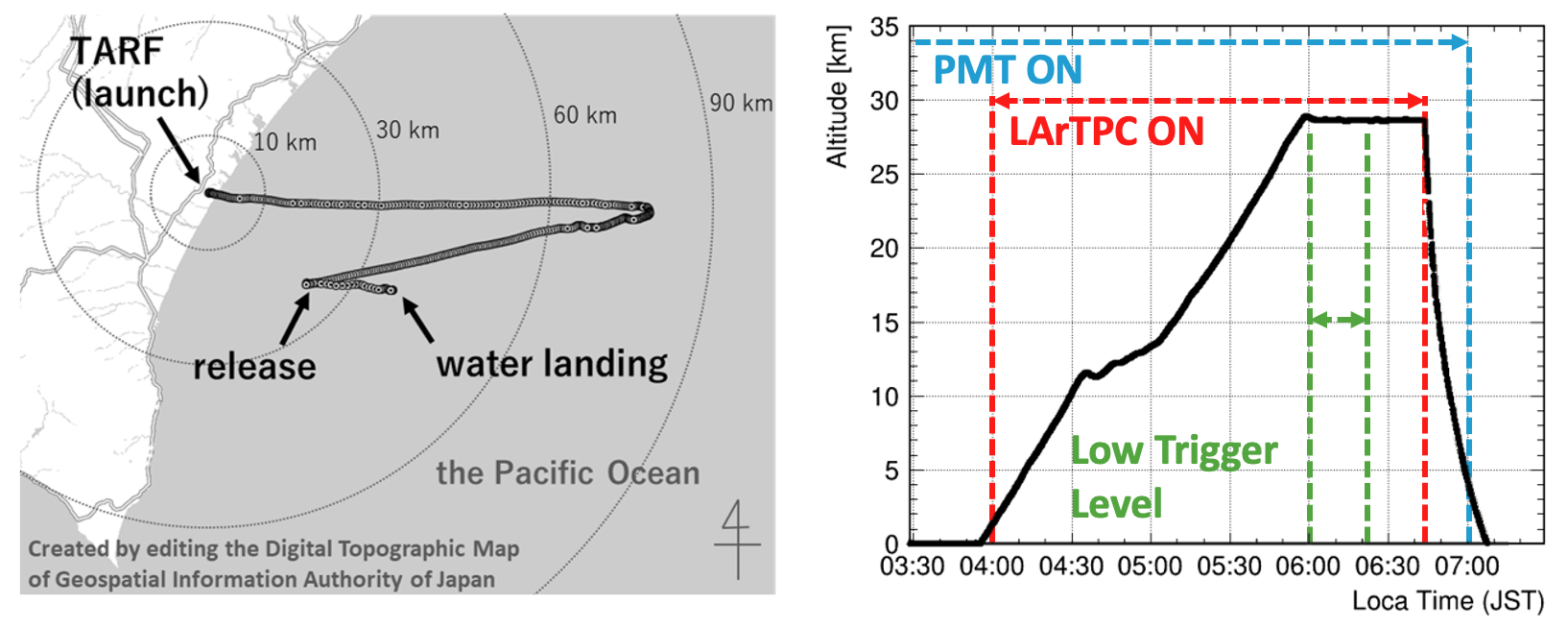}
  \end{center}
  \vspace{-0.4 cm}
  \caption[Flight pass and altitude for this mission]{Flight trajectory of B23-06 (left) and altitude (right) for this flight.}
  \label{fig:GPSmapBall}
\end{figure}

At 8:37~AM on July 26th, LAr was filled into the LAr vessel with enough amount for the flight on the following day. Until 12:00~AM, the LArTPC was operated to obtain cosmic muon events at the ground for calibration measures. At 2:28~AM on the day of the flight, the final operation check was completed and at 3:55~AM, the balloon was launched. Figure \ref{fig:GPSmapBall} shows the flight trajectory in the left panel and the flight altitude in the right plot with the detector operations for this mission. The GPS data were provided by JAXA/ISAS Balloon Group. As illustrated in the left panel of Figure \ref{fig:GPSmapBall}, the balloon follows a boomerang trajectory due to the directions of the jet stream winds at low altitudes being in the east direction and the winds at higher altitudes in the west direction. This boomerang flight enables a longer flight duration and recovery of the gondola closer to the land on the sea. The right plot of Figure \ref{fig:GPSmapBall} shows that the balloon was released at 03:55~AM on July 27th, 2023, after an ascent of 2 hours and 4 minutes, the balloon reached a maximum altitude of 28.9~km, and moved to a level flight at 05:59~AM for 44 minutes. At 06:43~AM, the gondola was released from the balloon and the gondola landed on the sea at 07:07~AM.

At 6:33~AM, before the gondola detachment, the solenoid valve (VF5) was opened to initiate the release of LAr before the release of the gondola from the balloon. The completion of the LAr release was confirmed at 07:00~AM during a soft descent facilitated by the parachute. Upon confirming the completion of the LAr release, a completion code was sent to the recovery team on the boat through iridium satellite communication. Subsequently, all power including the CPU, was immediately turned off. 

The environmental data, including pressure and temperature, along with PMT waveforms, were continuously recorded from the pre-launch phase until all power supplies were shut down.  The trigger level for the PMT signal was set to approximately 5 MeV of energy deposition inside the LArTPC active volume which was corresponding to a few cm of charged particle track. To prevent high voltage discharge within the LAr vessel, the high voltage to the cathode of the TPC was turned on 5 minutes after launch and turned off before the release of LAr. The trigger level was lowered for the first 20 minutes of the 40-minute level flight (500 keV for 10 minutes and 100 keV for another 10 minutes) to obtain gamma-ray events with low energy deposits. 

The payload safely landed on the sea at 07:07 AM and was recovered within a few minutes. Following recovery, no damage to the gondola, LAr container, TPC, or the pressurized vessel and its contents was observed.

\begin{figure}[htbp]
  \begin{center}
    \includegraphics[height=6.0cm]{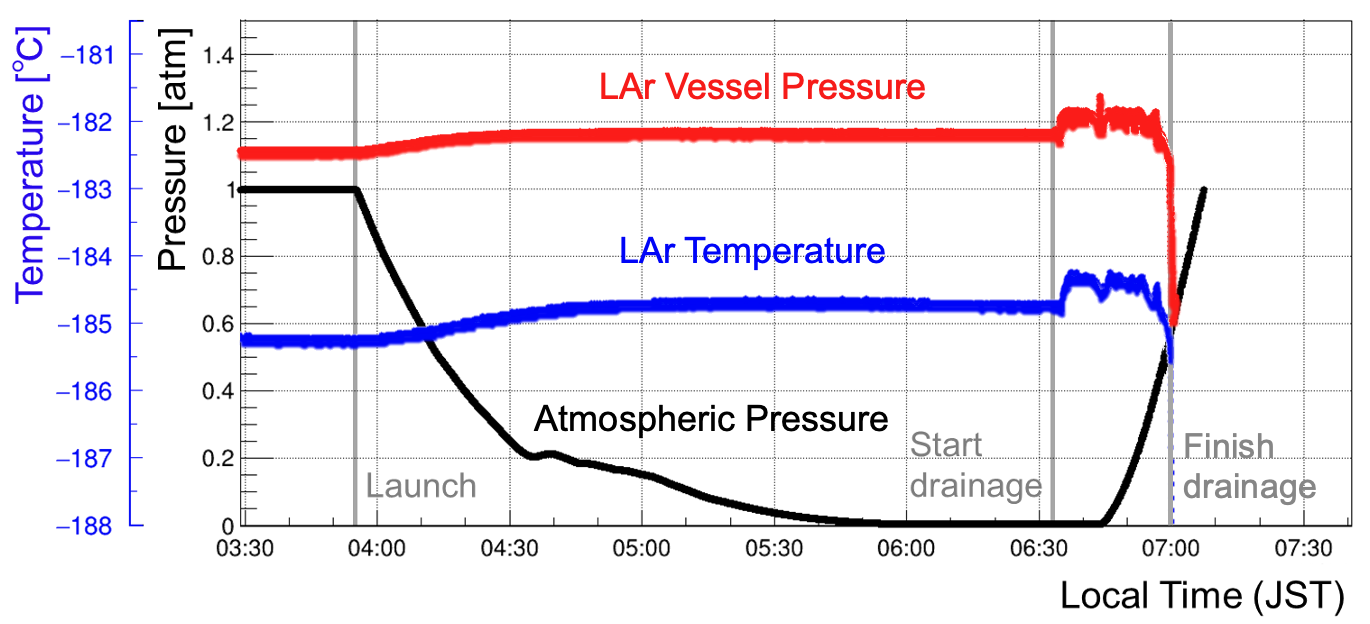}
  \end{center}
  \vspace{-0.6 cm}
  \caption[Changes in LAr vessel pressure, atmospheric pressure, and temperature during flight]{Changes in LAr vessel pressure (red), atmospheric pressure (black), and LAr temperature (blue) during flight.}
  \label{fig:Environment}
\end{figure}

Figure \ref{fig:Environment} displays the time variation of the pressure inside the LAr vessel (red line) and the temperature of the vessel bottom (blue line) alongside the atmospheric pressure (black line). The atmospheric pressure is represented as a value converted from altitude using the U.S. Standard Atmospheric Model \cite{ISA}. Throughout the flight, the pressure within the LAr vessel was consistently maintained within the expected range of 1.1 to 1.2~atm (due to the absolute pressure valve VF1). The thermometer, positioned below the liquid surface, monitors the LAr temperature, confirming that the argon remained in the liquid phase from the time of launch to the initiation of LAr evacuation. Hence, the internal pressure remained below the operating pressure of the differential pressure safety valve and rupture disk throughout the entire period from filling to recovery.

Next, our focus shifts to the LAr evacuation process: immediately after opening the solenoid valve (VF5), the vessel's pressure decreased, confirming the initiation of evacuation. However, within a few seconds, the liquid outlet closed, causing a subsequent pressure increase. As anticipated, the sudden decrease in atmospheric pressure below the triple point led to the rapid solidification of LAr near the evacuation piping outlet. Additionally, the LAr within the evacuation tube vaporizes, contributing to an elevated pressure inside the container compared to the pre-draining state. Due to the repetitive opening and closing of the solenoid valve (VF5) until the completion of liquid evacuation, the internal pressure experienced corresponding fluctuations. Following the separation of the payload from the balloon, the altitude decreased, resulting in an increase in atmospheric pressure to approximately the triple point. This change facilitated the smooth progression of the evacuation process. In this specific flight, LAr evacuation occurred after the altitude decreased. For future flights, enhancing the evacuation method such as by considering the installation of a heater in the evacuation line is recommended.

\section{LArTPC Data Analysis}\label{sec5}
\begin{figure}[htbp]
  \begin{center}
    \includegraphics[height=4.5cm]{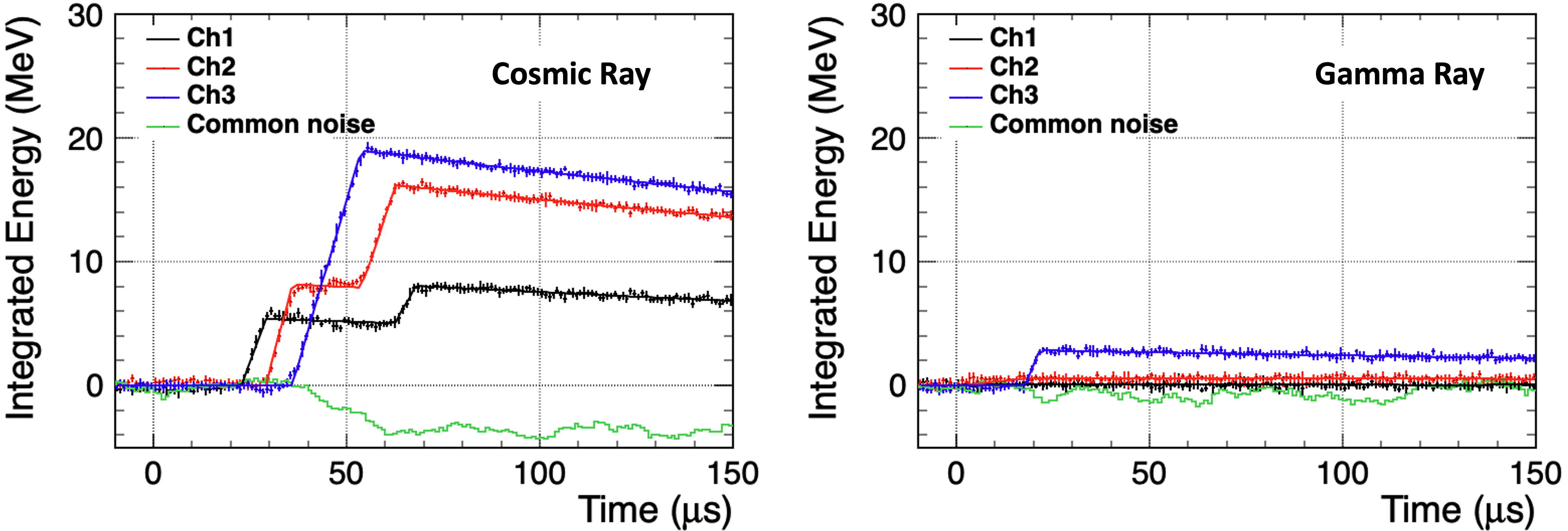}
  \end{center}
  \vspace{-0.4 cm}
  \caption[(Left) cosmic ray candidate (Right) gamma-ray candidate]{Example of cosmic ray (left) and gamma-ray (right) candidates obtained from the flight. (Black) outermost channel, (blue) innermost channel, (red) intermediate channel, (green) common noise.}
  \label{fig:Example}
\end{figure}

As mentioned in section \ref{sec4}, the LArTPC signals were successfully obtained throughout the flight. A total of around 0.5 million events were triggered by the PMT and detected by the LArTPC. The left and right plots in Figure \ref{fig:Example} illustrate a cosmic-ray penetration candidate and a gamma-ray Compton scattered candidate obtained during the level flight. The black, red, and blue lines represent the signal on each of the three anode channels. The green line corresponds to the common noise component (subtracted from each signal channel). For the cosmic-ray candidate, the signal on ch1 rises at 20~$\mu$s and 60~$\mu$s. Therefore, by considering the drift time of electrons, this could be understood as a cosmic ray entering the TPC from the side 2~cm below the anode and exiting the TPC from another side 6~cm below the anode. For the gamma-ray candidate, the signal on ch3 rises at 20~$\mu$s with no signals on the other channels. Therefore, this could be understood as a gamma-ray Compton scattering that occurred 2~cm below the anode, under the inner pad on the anode.  

\begin{figure}[htbp]
  \begin{center}
    \includegraphics[width=13cm]{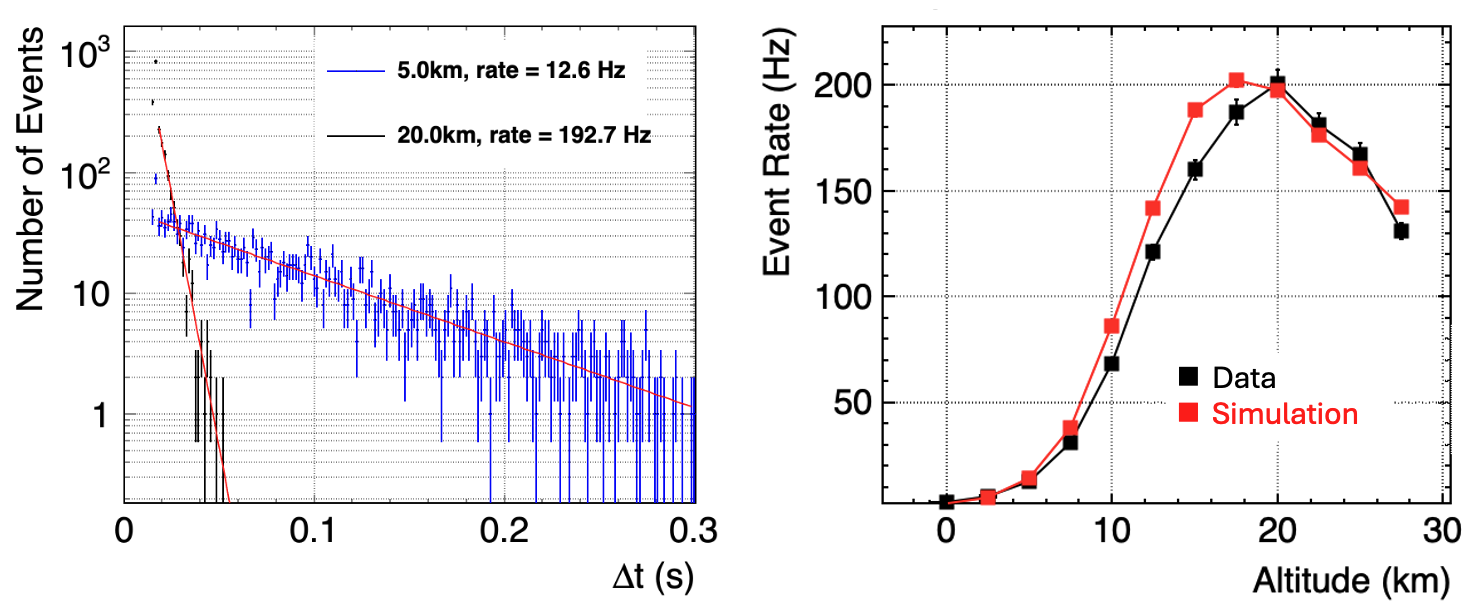}
  \end{center}
  \vspace{-0.4 cm}
  \caption[(Left) Distribution and fitting of time differences, (Right) Event rate as a function of altitude]{(Left) Time distribution between each event at 5~km altitude (blue dots), 20~km altitude (black dots), and the fitting results (red lines). (Right) Event rates at given altitudes are calculated from $\Delta$t distribution (black) Geant4 + EXPACS simulation result (red).}
  \label{fig:resR}
\end{figure}

The trigger rate was constrained by CPU performance, saturating at around 60~Hz. Therefore, the event rate $R$ at a given altitude was determined by fitting the distribution of the time difference ($\Delta t$) between each event to an exponential function: 
\begin{equation}
  \label{eq:RateFit}
  f(t) = A \times \mathrm{exp}(- R \times \Delta t),
\end{equation}
where $A$ is a scale factor and $R$ is the event rate. The left plot of Fig.~\ref{fig:resR} shows the distribution of the time difference at an altitude of 5~km (blue dots), 20~km (black dots), and the corresponding fitting results (red lines). The right plot of Figure~\ref{fig:resR} represents the rates calculated by exponential fitting shown in black. A Geant4-based simulation with input conditions from EXPACS \cite{EXPACS} providing angular and energy probability information of each particle at given altitudes was conducted as a comparison to the data. From the Geant4 simulation, the rate of events that deposited more than 5~MeV in the LArTPC is calculated and shown in red. As shown in Figure \ref{fig:resR}, for both data and the simulation, the event rate increases with the ascent of the balloon up to about 20~km, where the production of secondary particles from primary cosmic rays reaches its peak, known as the shower maximum, before decreasing with further altitude. 

\begin{figure}[htbp]
  \begin{center}
    \includegraphics[width=13cm]{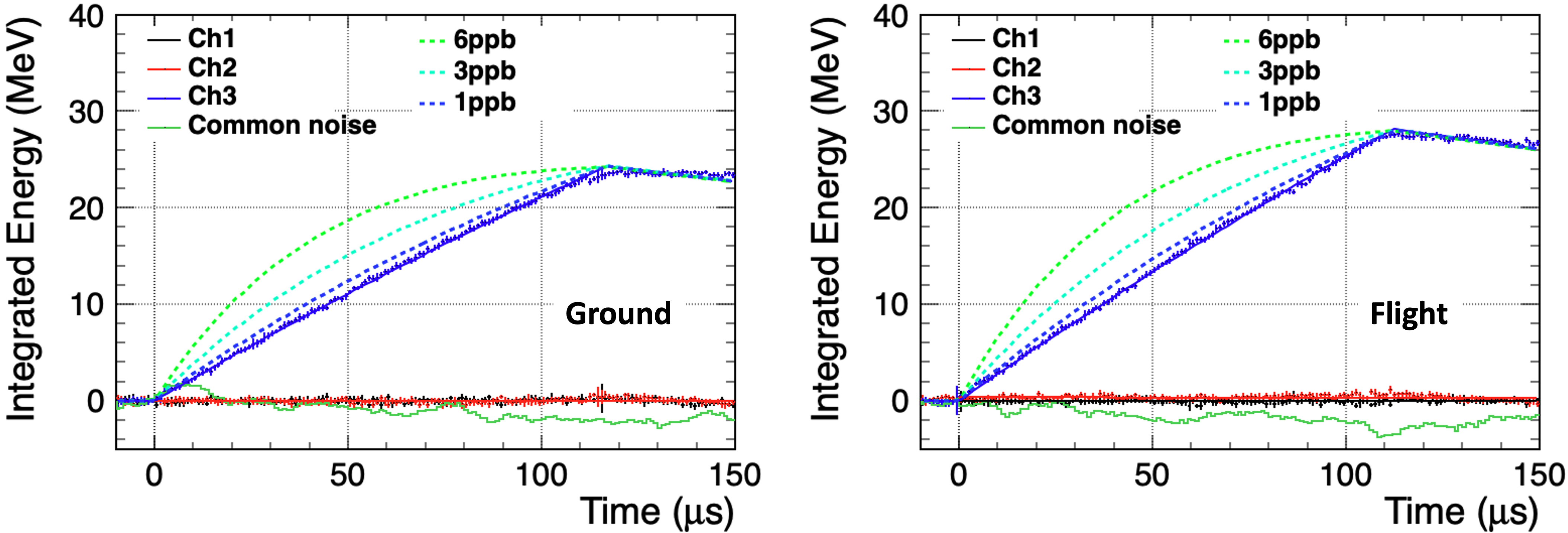}
  \end{center}
  \vspace{-0.4 cm}
  \caption{Waveforms of LArTPC signal from flight data (left) and ground data just before the launch (right). The colored dotted lines are predicted signals with different LAr purity.}
  \label{fig:purity}
\end{figure}

As mentioned in Section \ref{sec3.1}, LAr requires high purity for detecting charge signals on a LArTPC, and the purity needs to be maintained during the flight. In Figure \ref{fig:purity} the left and right plots are events obtained at level flight and ground, respectively. As shown in Figure \ref{fig:purity}, as the particle leaves only a signal on ch3 shown in blue, this event is a penetration event where a particle (most likely a Minimum Ionizing Particle) passes vertically through the TPC. The dotted lines represents fitted signal functions with the corresponding LAr impurities when the integrated energy of the signal is kept as a constant parameter. It can be seen that the purity was maintained in the flight as the purity was sub-ppb level before and during the flight.

\section{Summary and Future Prospects}\label{sec6} 

The GRAMS experiment is a next-generation balloon/satellite experiment aiming to detect cosmic MeV gamma-rays and cosmic antiparticles with a LArTPC. An engineering balloon flight with a small-scale LArTPC was launched from JAXA TARF in the summer of 2023 to establish a safe LAr handling system for balloon environments. The LAr was monitored and controlled safely during the 3-hour and 12-minute long flight. Also, cosmic ray data were obtained from the LArTPC along with gamma-ray data from the level flight. The analyses of the LArTPC data show that the LAr purity was maintained at sub-ppb level during the flight and from the rate calculation, the shower maximum was consistently found for both the data and simulation, resulting in a fully successful flight. 

Regarding future plans for GRAMS, an antiproton beam test at the J-PARC K1.8BR beamline is planned for winter 2024 to quantitatively understand the antiparticle capture reaction in LAr. Furthermore, a flight approved by the NASA/APRA program is planned between fall 2025 and spring 2026 in Arizona, USA. Finally, further development areas will be explored for the long-duration balloon flight in Antarctica using the full-size LArTPC detector.   

\addcontentsline{toc}{section}{Acknowledgements}
\section*{Acknowledgements}
\markboth{Acknowledgements}{Acknowledgements}

We thank the Scientific Ballooning (DAIKIKYU) Research and Operation Group, ISAS, JAXA for the professional support of the eGRAMS flight. We are grateful for the support of Prof. K.~Ishimura for the gondola design support. Also, we are grateful for the support of Prof. Y.~Makida for providing the absolute pressure valve and for consultations on the safety design. This work was supported by the JSPS Grant-in-Aid for Scientific Research (A) (22H00133), (B) (22H01252), and Challenging Research (Pioneering) (22K18277) in Japan. We also acknowledge support by NASA under Award No.22-APRA22-0128 (80NSSC23K1661) and the Alfred P. Sloan Foundation in the USA.


\let\doi\relax 

\bibliographystyle{plain} 
\bibliography{main}

\end{document}